\documentstyle[aps,psfig,floats,graphicx]{revtex}

\begin{document}
\newcommand{\be}{\begin{eqnarray}}
\newcommand{\ee}{\end{eqnarray}}
\newcommand\del{\partial}
\newcommand\nn{\nonumber}
\newcommand{\Tr}{{\rm Tr}}

\wideabs{
\begin{flushright}
NORDITA-2004-17 HE \\
SUNY-NTG-04/01 
\end{flushright}

\title  {Supersymmetric Quenching of the Toda Lattice Equation}

\author {K. Splittorff$^1$ and J.J.M. Verbaarschot$^2$}

\address{$^1$ Nordita, Blegdamsvej 17, DK-2100, Copenhagen {\O},
  Denmark \\ $^2$ Department of Physics and Astronomy, SUNY, Stony Brook, New
  York 11794, USA\\email:split@alf.nbi.dk and
  verbaarschot@tonic.physics.sunysb.edu}

\date   {\today}
\maketitle
\begin  {abstract}
The average of the ratio of powers 
of the spectral determinants of the Dirac operator in the 
$\epsilon$-regime of QCD is shown to satisfy a Toda lattice equation. 
The quenched limit of this Toda lattice equation is obtained using the
supersymmetric method. This super symmetric approach is then shown to be
equivalent to taking the replica limit of the Toda lattice equation. 
Among other, the factorization
of the microscopic spectral
correlation functions of the QCD Dirac operator into fermionic and bosonic 
partition functions follows naturally from both approaches. 
While the replica approach relies on an analytic
continuation in the number of flavors no such assumptions are made in the
present approach where the numbers 
 of flavors in the Toda lattice equation 
are strictly integer.
\end {abstract}
}

\widetext
\onecolumn

\vspace{1cm}

\section{Introduction}
 
A quenched theory is one in which a determinant has been removed 
from weight in the partition
function. Removing a determinant from the partition function 
does not necessarily lead to an unphysical theory. The 
determinant can act as a source term for Green's functions,
and, once a derivative of the 
partition function has been taken, this source is removed. 
Two widely used analytical tools to remove determinants are the
replica trick and the supersymmetric method. For the replica trick,  
one introduces $n$ identical copies of the determinant, 
performs the appropriate
derivatives, and finally takes the limit $n\to 0$. For the supersymmetric 
method, a ratio of two determinants with different values of the sources
is introduced in the partition function. Derivatives are
performed with respect to the sources and the determinants are removed
by setting the sources equal. 
This paper deals with the equivalence of these two approaches.

Originally, both the replica trick \cite{EA} and the supersymmetric method
\cite{Efetov} where invented for the study of disordered 
systems.
Applications of the replica trick soon
proved to be delicate, see eg. \cite{Mezard}, the trouble being in the choice
of the dominant saddle points and the analytic continuation in the replica
index \cite{VZ}.  
The supersymmetric method \cite{Efetov} does not rely on such analytic
continuation, and never faces this problem. At the perturbative level,
one only finds a polynomial dependence on the number of replicas and
both approaches yield identical results 
(see for example \cite{VZ,wegner-super,kim}).
The replica trick has made its mark in diverse areas of 
theoretical physics. Some examples are the proof of the linked
cluster theorem (see \cite{negele-book}) and the study of vacua in
supersymmetric gauge theories \cite{deboer}.

In disordered systems, which also includes the Dirac operator in
a gauge field distributed according to the Yang-Mills action, the
standard approach both for the replica trick and the supersymmetric
method is to rewrite the bosonic and/or fermionic integrals in terms
of a non-linear $\sigma$-model. Such a $\sigma$-model describes the
Goldstone modes associated with the  spontaneous breaking of the
global symmetries of the original model. Both its symmetry breaking
term and its kinetic term are determined by the pattern of global
symmetry breaking and Lorentz invariance. In the theory of disordered
systems, these Goldstone modes are known as diffusons whereas
in QCD they are known as the pseudoscalar mesons (for reviews see 
\cite{Efetov,L}). For QCD, the 
$\sigma$-model is known as a chiral Lagrangian and has been studied
extensively in relation to the low-energy properties of the strong
interactions \cite{GL}.

In a parameter domain where the kinetic term is important, in most cases,
only a 
perturbative treatment of the theory is feasible. However, in a
domain where the kinetic term can be ignored, which is know as
the ergodic domain or the $\epsilon$ domain, the $\sigma$-model 
becomes zero dimensional and can be solved exactly in many cases.
The zero dimensional $\sigma$-models are equivalent to random matrix
theories \cite{Efetov,VWZ,class,OTV,DOTV}. Using this connection, 
the problems with the replica trick where shown to be of a more serious
nature \cite{critique}.
This lead to the general
consensus that the replica trick was only reliable for perturbative
calculations and could generally not be used to obtain exact 
nonperturbative 
results.  A first hint that nonperturbative results could
be obtained from the replica trick came from \cite{KM,lerner} 
where the oscillatory factors in the
asymptotic expansion of the two-point correlations function were reproduced.
However, only
in cases where the perturbative expansion terminates \cite{zirn}, exact
analytical results could be obtained, and, as was pointed out in \cite{zirn},
a systematic and exact replica method remained illusive in these works.  
Recently, replica methods were introduced that went beyond these limitations 
\cite{kanzieper02,SplitVerb1,SplitVerb2,kanzieper03}. 
Most significantly,
exact analytical results were obtained in cases where other methods
have failed.
While the formulation of
the replica trick introduced in \cite{SplitVerb1} has been 
shown to be widely applicable, it still relies on an assumed    
analytical continuation in the replica index. The central purpose of this
paper is to 
justify this analytical continuation.

The new development which allowed for an exact analytic evaluation of the
spectral correlation functions is the link between zero dimensional
$\sigma$-models and an exactly solvable system know as the Toda lattice. 
This system is one dimensional and consists of massive particles interacting 
through an exponential potential \cite{todabook}.
The Hamilton equations of motion for the position of the masses as a function
of time are known as the Toda lattice equations. 
The system is integrable, and its solution for appropriate boundary
conditions is given by the partition function of a zero dimensional
$\sigma$-model.

Previously, we have shown \cite{SplitVerb1,SplitVerb2} that the
replica limit of the Toda lattice equation not only gives the exact analytic
results but also explains the factorization 
of the universal microscopic correlation functions into a product of a
bosonic and a fermionic partition function. 
Though this factorization suggests a direct relationship with the
supersymmetric generating functionals, it has remained a miracle why
the final result  factorizes within the supersymmetric method.
In this paper we show this factorization  based on explicit expressions
for the ratio of different powers of spectral determinants \cite{Yan}.
The key observation is that this family of partition functions
also satisfies Toda lattice equations. Quenching
these Toda equations as in the supersymmetric method, that is setting 
bosonic and fermionic masses equal, automatically reveals the
factorization of correlation functions. As stressed above,  
the replica limit of the Toda lattice equation relies on an analytic
continuation in the number of flavors from integer values to real
values. Although the replica limit of the Toda lattice equation, in  cases 
that have been studied up to now, has given the correct analytic result, 
no definite arguments have been given that 
the usual problems of the replica limit
do not manifest themselves in this approach. As we will 
show in this paper the replica limit of the Toda lattice equation is
equivalent to quenching the Toda lattice equation of a family of 
supersymmetric partition functions with a strictly integer number of
bosonic and fermionic flavors. 
This result
directly justifies the interpretation of the partition function with
zero flavors in the usual Toda lattice equation as the supersymmetric
partition function.

Throughout this paper we consider the zero dimensional limit of the 
$\sigma$-model relevant for QCD. In the random matrix formulation this
corresponds to the chiral unitary ensemble (chUE).
We expect that similar results can be obtained for the unitary
ensemble.     

This paper is organized as follows. First we describe the replica method
and the supersymmetric method. Then, in section \ref{sec:TodaEqs},
 we discuss Toda lattice
equations relevant for the one-point function and the two-point
function. In section \ref{sec:susyQuench} the quenching of 
the Toda lattice equations is discussed.
This automatically yields the quenched
correlation functions in the factorized form. 
The equivalence of the replica method and 
the supersymmetric approach is discussed 
in section \ref{sec:equiv}.
Finally, in section \ref{sec:conc}, we summarize our results and comment on
possible extensions of our work. In two appendices we derive the Toda lattice 
equations for the supersymmetric partition functions.

\section{Definitions of Replica and Supersymmetric quenching}

In this section we set our notation and introduce the supersymmetric method 
and the replica method. We illustrate the general definitions
with generating functions for the microscopic limit of
the one-point functions of the chUE. 

\vspace{5mm}

The essential ingredient in the partition functions are the fermionic and
bosonic determinants which will be written explicitly  whereas all
other parts of the weight are included 
in the average denoted by $\langle \ldots\rangle$.
The fermionic replicated partition function is defined by
\be
\label{ZNf}
Z_{N_f}(x) = \left\langle {\det}^{N_f}(D+x) \right\rangle,
\ee
where $D$ is an operator, the mass $x$ is a number, and, as just 
mentioned,
$\left\langle\ldots \right\rangle$ denotes the average for the
specific theory (note that  $\del_x\left\langle 1 \right\rangle=0$).
Analogously, the bosonic replicated partition function is given by 
\be
Z_{-N_b}(y) = \left\langle {\det}^{-N_b}(D+y) \right\rangle.
\ee
The replica trick for  removing the determinants is to take the limit
$N_f\to0$ or $N_b\to0$. 

The supersymmetric partition function is defined by the average ratio of
powers of the spectral determinants
\be
Z_{N_f,-N_b}(x|y) = \left\langle {\det}^{N_f}(D+x){\det}^{-N_b}(D+y) \right\rangle
\ee
which can be removed by taking the limit $y\to x$. For example, 
\be
\lim_{y\to x} Z_{4,-2}(x|y) =  Z_2(x).
\ee

Obviously, it is not particularly interesting to introduce and
remove determinants unless we differentiate with 
respect to the masses first.
The average resolvent of the operator $D$ is defined as
\be
G(x)  \equiv  \left\langle {\rm Tr} \frac 1{D+x} \right\rangle.
\ee
In {\sl the fermionic replica limit} it is given by
\be
G(x) = \lim_{N_f\to0} \frac{1}{N_f} \del_x \log Z_{N_f}(x) ,
\ee
in {\sl the bosonic replica limit} we have that
\be
G(x) = \lim_{N_b\to0} \frac{1}{-N_b} \del_x \log Z_{-N_b}(x) ,
\ee
and {\sl the supersymmetric method} is based on the identity 
\be
G(x) = \lim_{y\to x} \del_x \log Z_{1,-1}(x|y) .
\ee
 
While the average in eq. (\ref{ZNf}) appears to be well defined for any real
value of $N_f$ it is often not possible to evaluate the average unless $N_f$
is an integer. For this reason, taking the limit $N_f\to0$ involves an
analytic continuation in $N_f$. 

In this paper we focus on the topological trivial sector of the QCD 
partition function which is the average of a fermionic
determinant 
\be
{\cal Z}_{N_f}(\{m_f\}) = \int[{\rm d}A]_0 [\prod_{f=1}^{N_f}
\det(i\gamma_\mu D_\mu-m_f)]\,e^{-S_{\rm YM}(A)}
\ee
over gauge field configurations with trivial topology weighted by the
Yang-Mills action $S_{\rm YM}$. The quark masses $\{m_f\}\equiv
m_1,\ldots,m_{N_f}$ may take different or identical values. 
In particular, we are interested in spectral
correlation functions of the Dirac operator $\gamma_\mu D_\mu$ in the phase
where chiral symmetry is spontaneously broken. The order parameter for the
spontaneous breakdown of chiral symmetry is the chiral condensate denoted by 
$\langle\bar{\psi}\psi\rangle$. At low energies, the effective
degrees of freedom are the Goldstone modes associated with this spontaneously
broken symmetry, and the effective theory is a $\sigma$-model
\cite{Weinberg}. The  zero dimensional limit of this $\sigma$-model allows for
an analytical evaluation of the correlation functions on
microscopic scales. The microscopic scale is set by the inverse of
$V\langle\bar{\psi}\psi\rangle$, where $V$ is volume of the system. 
The domain in which the zero momentum modes of
this effective theory decouple from the nonzero modes 
is known as the $\epsilon$-regime of QCD \cite{GL}. In
the topologically trivial sector, $\nu=0$, the fermionic partition
functions are given by the unitary, and hence compact, integral
\cite{GLeps,LS,SV}  
\be
Z_{N_f}(\{x_f\}) = \int_{U \in U(N_f) } \hspace{-2mm} dU e^{\frac 12
{\rm  Tr}[M^\dagger U +M U^\dagger]},
\label{Zf}
\ee
where $M\equiv{\rm diag}(x_1,\cdots,x_{N_f})\equiv{\rm
  diag}(\langle\bar{\psi}\psi\rangle V
m_1,\cdots,\langle\bar{\psi}\psi\rangle V m_{N_f})$ is the rescaled mass
matrix. 
The bosonic partition functions are given by a non-compact integral over
positive definite matrices \cite{OTV,DV1}
\be
Z_{-N_b}(\{y_b\}) = 
\int_{Q \in Gl(N_b)/U(N_b) } \hspace{-12mm} dQ 
e^{-\frac 12{\rm  Tr}[M^\dagger Q +M Q^{-1}]},
\label{Zb}
\ee
where the dependence on the bosonic quark masses, the chiral condensate, and
the volume is through the products $y_b\equiv m_b\langle\bar{\psi}\psi\rangle
V$ appearing in the rescaled quark mass matrix 
$M\equiv{\rm  diag}(y_1,\cdots,y_{N_b})$. The supersymmetric
partition function of QCD in the $\epsilon$-regime is given by an integral
over a supergroup \cite{class,OTV} 
\be
Z_{N_f,-N_b}(\{x_f\}|\{y_b\}) = \int_{Q \in \hat{Gl}(N_f|N_b)} \hspace{-12mm} dQ 
e^{\frac 12{\rm  Str}[M^\dagger Q +M Q^{-1}]},
\label{Zsusy}
\ee
where $M={\rm diag}(x_1,\cdots,x_{N_f},y_1,\cdots,y_{N_b})$ is the rescaled
mass matrix. The boson-boson block of $\hat{Gl}(N_f|N_b)$ is
$Gl(N_b)/U(N_b)$ and the fermion-fermion block of $\hat{Gl}(N_f|N_b)$ is 
$U(N_f)$. 
All three integrals are fixed uniquely by the symmetries.

Much has been learned about these partition functions over the past few 
years. Explicit expressions in terms of determinants of Bessel
functions are known in all cases 
\cite{OTV,SplitVerb1,Yan,LS,DV1,tan,dijkgraaf,mmm,FS,FSgen,FA,jsv,Tilo,baba}. 
For $Z_{N_f}$ and $Z_{-N_b}$ the expressions have been derived directly from
the group integrals (\ref{Zf}) and (\ref{Zb}) while the derivation of the 
general expression for $Z_{N_f,-N_b}$ \cite{Yan,FS,FSgen,FA} 
goes through the random matrix
representation. For all the partition functions $N_f$ and
$N_b$ must take integer values and the analytic continuation in $N_f$
and $N_b$ is not uniquely defined. 

As mentioned in the introduction, the latest developments include 
a link between such matrix integrals and an exactly solvable one
dimensional system known as the Toda lattice  \cite{dijkgraaf,mmm,forrester}. 
The Hamiltonian equation 
of motion of the Toda lattice is known as the Toda lattice equation. 
The remarkable link is
that the QCD partition functions also obey such Toda lattice
equation. This equation connects partition functions with different
numbers of flavors, much in the same way as recursion relations
for orthogonal polynomials connect polynomials with different indices.
It turns 
out that the Toda lattice equation for fermionic flavors is
also valid for bosonic flavors.
Both in the bosonic case and the fermionic case the Toda lattice equation
terminates for zero flavors. Based on the correct analytical result 
for the replica limit of these hierarchies, it was conjectured in 
\cite{SplitVerb1} that both hierarchies are connected by the supersymmetric
partition function. 
This assertion is proved in this paper starting from 
the observation that the supersymmetric partition
functions also obey a Toda lattice equation.

\section{The Toda Lattice Equations for the Supersymmetric Partition Function}
\label{sec:TodaEqs}

The Toda lattice equations considered in the literature (see for instance
\cite{forrester,Litt,Kharchev})   
connect partition functions with a different number of fermionic 
flavors or a different number of bosonic flavors. That is, the Toda
lattice equations deal with semi-infinite hierarchies with either positive
or negative index. In \cite{SplitVerb1,SplitVerb2} it was found that
these two semi-infinite hierarchies are connected.  
The connection was realized as the
replica limit of the index in either the positive or the 
negative semi-infinite hierarchy. 
Here, we consider Toda lattice equations for the 
partition functions with both fermions and bosons, i.e. 
super-symmetric partition functions. 
Referring to the graded symmetry of the supersymmetric partition functions
we call these graded Toda lattice equations.  
In the next section we will derive 
the quenched resolvents from these graded Toda lattice equations.

\subsection{A graded Toda Lattice Equation Affecting One Flavor}

Starting from the expression for the partition function (\ref{Zsusy}) 
with $N_f$ fermionic flavors and $N_b$ bosonic flavors
\cite{FS,FSgen,FA,SplitVerb1}\footnote{{\sl Notation:} The vertical bar in 
the argument of $Z$ separates 
the masses $\{x_f\}\equiv  x_1,\ldots,x_{N_f}$ of the fermionic quarks
from the masses $\{y_b\}\equiv y_1,\ldots,y_{N_b}$ of the bosonic quarks. For
degenerate masses we either repeat the mass in the argument of $Z$
according to the degeneracy or put an additional subscript on $Z$ indicating
the degeneracy, eg. $Z_{3,1,-2}(x_1,x_2|y)\equiv
Z_{4,-2}(x_1,x_1,x_1,x_2|y,y)$. For bosonic masses, the subscritpt of 
$Z$ that denotes the number of flavors will contain and explicit minus sign.
If all flavors are of the same kind the vertical bar will be omitted.}
\be
\label{Z-nm}
Z_{N_f,-N_b}(\{x_f\}|\{y_b\})\!=\!\frac{\det[z^{j-1}_i{\cal
J}_{j-1}(z_i)]_{i,j=1,..,N_f+N_b}} 
{\prod_{j>i=1}^{N_f}(x_j^2-x_i^2)\prod_{j>i=1}^{N_b}(y_j^2-y_i^2)},  
\ee
where $z_i=x_i$ for $i=1,\ldots,N_f$,\, $z_{N_f+i}=y_i$ for
$i=1,\ldots,N_b$, \,${\cal J}_{j-1}(z_i)\equiv I_{j-1}(x_i)$ for
$i=1,\ldots,N_f$, and ${\cal J}_{j-1}(z_{N_f+i})\equiv (-1)^{j-1}K_{j-1}(y_i)$ for $i=1,\ldots,N_b$, we prove in \ref{App:Toda1PF} that the partition function
(\ref{Zsusy}) satisfies the graded Toda lattice relation  
\be
\label{Toda1PF}
\delta_{N_f+N_b}\delta_{x_i}\log Z_{N_f,-N_b}(\{x_f\}|\{y_b\})
= 2 n x_i^2\frac{ Z_{N_f+1,-N_b}(x_i,\{x_f\}|\{y_b\})
Z_{N_f-1,-N_b}(\{x_f\}_{f\neq i}|\{y_b\})}{[Z_{N_f,-N_b}(\{x_k\}|\{y_b\})]^2}.
\ee
Here, $n$ is the number of fermionic flavors with mass $x_i$ in 
$Z_{N_f,-N_b}(\{x_f\}|\{y_b\})$. The derivation $\delta_x$ is defined by
\be
\delta_x \equiv x\del_x
\ee
and 
\be 
\label{def-deltaNfNb}
\delta_{N_f+N_b} \equiv \sum_{f=1}^{N_f}\delta_{x_f}+\sum_{b=1}^{N_b}\delta_{y_b}.
\ee
For degenerate masses the sum only extends over the different masses.

This is the generalization of the Toda lattice equation for fermions 
obtained 
in \cite{Kharchev}, to include both fermions and bosons. Below, in
section \ref{subsec:1pf}, we quench this graded Toda lattice equation as in the
supersymmetric method and show that the factorization of the quenched
resolvent follows naturally.

\subsection{A graded Toda Lattice Equation Affecting Two Flavors}

The key ingredient in understanding the factorization property
\cite{SplitVerb2} of the
two-point correlation function within the supersymmetric approach is 
a graded Toda lattice equation where two indices are raised and lowered
simultaneously. In \cite{SplitVerb2} it was shown that for $m$ degenerate
fermionic flavors with mass $x_1$ and $n$ degenerate fermionic flavors
with mass $x_2$, the partition function (\ref{Zf}) satisfies the Toda lattice
equation ($m$ and $n$ are positive integers)
\be
\delta_{x_1}\delta_{x_2} \log Z_{m,n}(x_1,x_2) = 4 m n x_1^2 x_2^2 
\frac{Z_{m+1,n+1}(x_1,x_2)Z_{m-1,n-1}(x_1,x_2)}{[Z_{m,n}(x_1,x_2)]^2}.
\ee
Using the explicit expressions for the partition function given in
(\ref{Z-nm}) we have shown that
\be
\label{TheToda2f}
 \delta_{x_1}\delta_{x_2} \log Z_{2,-2}(x_1, x_2|y_1, y_2) 
 = 4   
x_1^2 x_2^2 \frac{ Z_{4,-2}(x_1,x_1,x_2,x_2|y_1, y_2)
Z_{-2}(y_1,y_2)}
{[Z_{2,-2}(x_1,x_2|y_1,y_2))]^2}
\ee
and a similar equation for other small numerical values
of the number of flavors (see \ref{App:Toda2PF}).

We conjecture that these results can be
generalized to  any integer number of fermions, 
$N_f$, and any integer number of bosons, $N_b$ (further support for this
conjecture will be given in \ref{App:Toda2PF}):
\be
\label{TheToda2f-conj}
 \delta_{x_i}\delta_{x_j} \log Z_{N_f,-N_b}(\{x_f\}|\{y_b\}) 
 = 4 nm 
x_i^2 x_j^2 \frac{ Z_{N_f+2,-N_b}(x_i,x_j,\{x_f\}|\{y_b\})
Z_{N_f-2,-N_b}(\{x_f\}_{f\neq i,j}|\{y_b\})}
{[Z_{N_f,-N_b}(\{x_f\}|\{y_b\})]^2}, 
\ee
where $m$ and $n$ are the number of flavors with mass $x_i$ and $x_j$
in $Z_{N_f,-N_b}(\{x_f\}|\{y_b\})$, respectively. 
A similar equation holds for differentiation with respect to 
 two bosonic masses.
In that case the fermionic flavors are passive.
Furthermore, we have shown for a number of values of $N_f$ and $N_b$
 (see \ref{App:Toda2PF}) that a corresponding equation is  valid 
when we differentiate with respect to one fermionic and one bosonic
mass. Without proof we assert that the general result in this
case is given by
\be 
\label{TheToda1f1b}
\delta_{x_i}\delta_{y_j} \log Z_{N_f,-N_b}(\{x_f\}|\{y_b\}) 
 = -4nm x_i^2 y_j^2 \frac{ Z_{N_f+1,-N_b+1}(x_i,\{x_f\}|\{y_b\}_{b\neq j})
Z_{N_f-1,-N_b-1}(\{x_f\}_{f\neq i}|y_j,\{y_b\})}
{[Z_{N_f,-N_b}(\{x_f\}|\{y_b\})]^2}, 
\ee 
where $m$ and $n$ are the number of flavors with mass $x_i$ and $y_j$
in $Z_{N_f,-N_b}(\{x_f\}|\{y_b\})$.
These graded Toda lattice equations go beyond previous results
\cite{SplitVerb2} in two respects:
{\sl 1)} they have been extended to supersymmetric partition functions, and
{\sl 2)} the masses in these equations can be either 
         degenerate or non degenerate.

\section{Supersymmetric Quenching of the graded Toda Lattice Equation}
\label{sec:susyQuench}

In the supersymmetric approach the partially quenched  or fully quenched 
correlation functions are obtained by differentiating before taking the limit
of equal fermionic and bosonic masses. In this section we show how the graded 
Toda lattice equations of the previous section automatically give the
quenched one-point and two-point functions. 
The results are in exact agreement with
expressions obtained previously. The important new result of this section 
is that the factorization of correlation functions into products of
partition functions follows naturally without assuming an analytical
continuation in the number of flavors.

\subsection{The One Point Function}
\label{subsec:1pf}

In this section we derive the quenched spectral one point function by
quenching the graded Toda lattice equation (\ref{Toda1PF}) using the
supersymmetric 
method. For this it is sufficient to have one bosonic and one fermionic
flavor, and the graded Toda lattice equation then reads
\be
\label{Toda1PF:1-1}
\delta_x(\delta_x+\delta_y)\log Z_{1,-1}(x|y)
= 2 x^2\frac{ Z_{2,-1}(x,x|y)Z_{-1}(y)}{[Z_{1,-1}(x|y)]^2}.
\ee
the limit $y\to x$ of the l.h.s. of this equation 
is given by 
\be
\label{22}
\lim_{y\to x}\delta_x(\delta_x+\delta_y)\log Z_{1,-1}(x|y)
 & = &
\lim_{y\to x} \delta_x \left [x\left\langle\Tr\frac{1}{D+x}\right\rangle_{x,y}
-y\left\langle\Tr\frac{1}{D+y} \right\rangle_{x,y} \right]
\nn \\ 
 & = &\delta_x x G(x).
\ee 
Here, we have used the notation 
\be
\left\langle\ldots\right\rangle_{x,y}=
\left\langle\ldots\frac{\det(D+x)}{\det(D+y)}\right\rangle. 
\ee
Taking also the limit $y\to x$ of the r.h.s. of (\ref{Toda1PF:1-1})
we find
\be
\label{1pfFacto} 
\delta_x x G(x)=2 x^2 Z_{1}(x)Z_{-1}(x).
\ee
This result is in complete agreement with known results for the
resolvent and the partition functions. For completeness we mention that
they are given by
\be
G(x)= x(K_0(x)I_0(x)+K_{-1}(x)I_{1}(x)),\qquad Z_1(x) =I_0(x),\qquad Z_{-1}(x)
= K_0(x).
\ee
Clearly, this is a rather cumbersome way of deriving the resolvent from
$Z_{1,-1}$. 
Let us therefore again stress that the important new result here is that 
the factorization 
in eq. (\ref{1pfFacto}) is proved on the basis of the
Toda lattice hierarchy for the supersymmetric 
partition functions. We 
also
stress that this approach does 
not rely on an analytic continuation
in the number of flavors. 

Following the same strategy as above one can also obtain the
partially quenched resolvents. As we seek to stress a matter of principle 
rather than reproducing existing results we instead turn to the two-point
function.       

\subsection{The Two-Point Function}
\label{subsec:TodaLattEq2pf}

The two-point disconnected susceptibility contains two derivatives just
as in the Toda lattice equation. This property simplifies the
derivation of two point-functions from the Toda
lattice  equation for ratios of spectral determinants.

The fully quenched disconnected susceptibility in the chiral unitary ensemble
with identical arguments is defined as
\be
\chi(x,x)\equiv-\lim_{y\to x}\del_{x}\del_{y}\log Z_{1,-1}(x|y). 
\ee
It follows from
(\ref{TheToda1f1b}) for $N_f=N_b=1$ at equal values of the arguments that
\be 
\chi(x,x)\equiv-\lim_{y\to x}\del_{x}\del_{y}\log Z_{1,-1}(x|y) = 4 x^2
Z_2(x,x)Z_{-2}(x,x). 
\ee 
In order to get the quenched disconnected susceptibility for different
values of the arguments,
we start from
the partition function describing two fermions and two bosons all with
different masses. We now have a choice: we can differentiate with
respect to two fermionic masses, with respect to two bosonic masses,
or with respect to one fermionic mass and one bosonic mass, before
pairing up the fermion and boson masses. If we
choose the first option and use (\ref{TheToda2f}) we find
\be
\label{twentyeight}
\chi(x_1,y_1)&\equiv& -\lim_{y_1\to x_1,y_2\to x_2}\del_{x_1}\del_{x_2}\log
Z_{2,-2}(x_1,x_2|y_1,y_2) \\
& = & 
\lim_{y_1\to x_1,y_2\to x_2} 4 x_1x_2
\frac{Z_{4,-2}(x_1,x_1,x_2,x_2|y_1,y_2)Z_{-2}(y_1,y_2)}
{[Z_{2,-2}(x_1,x_2|y_1,y_2)]^2} \nn\\
& = & 4 x_1x_2 Z_2(x_1,x_2)Z_{-2}(x_1,x_2).\nn 
\ee
The derivation when differentiating with respect to the two bosonic masses
is analogous and leads to exactly the same quenched correlation function. 
If we choose to differentiate with respect to one
fermionic mass and one bosonic mass and 
use (\ref{TheToda1f1b}) for $N_f=N_b=2$, 
\be
\del_{x_2}\del_{y_1}\log
Z_{2,-2}(x_1,x_2|y_1,y_2) 
& = & -4 x_2y_1
\frac{Z_{3,-1}(x_1,x_2,x_2|y_2)Z_{1,-3}(x_1|y_1,y_1,y_2)}
{[Z_{2,-2}(x_1,x_2|y_1,y_2)]^2}, 
\ee
(which we have also verified explicitly) we obtain
\be 
\chi(x_1,x_2)&\equiv& -\lim_{y_1\to x_1,y_2\to x_2}\del_{x_2}\del_{y_1}\log
Z_{2,-2}(x_1,x_2|y_1,y_2) \\
& = & 
\lim_{y_1\to x_1,y_2\to x_2} 4 x_2y_1
\frac{Z_{3,-1}(x_1,x_2,x_2|y_2)Z_{1,-3}(x_1|y_1,y_1,y_2)}
{[Z_{2,-2}(x_1,x_2|y_1,y_2)]^2} \nn\\
& = & 4 x_1 x_2 Z_2(x_1,x_2)Z_{-2}(x_1,x_2).\nn 
\ee
This result coincides with (\ref{twentyeight}) and 
is in agreement with the result obtained from a
supersymmetric calculation \cite{TV-two} and with the result obtained
from the replica limit of the Toda lattice equation of the fermionic 
hierarchy \cite{SplitVerb2}.
Most importantly, it is clear that the factorization 
property of the quenched two-point function in the supersymmetric 
approach is a
direct consequence of the structure in the graded Toda lattice equation.
No assumptions on the analytic continuation in the flavor index
have to be made in this case.

\section{Equivalence of Replica and Supersymmetric Quenching}
\label{sec:equiv}

In this section we show 
that the replica limit of the Toda lattice
equation necessarily gives the same answer as the supersymmetrically
quenched Toda lattice equation. This directly justifies the central 
assumption of\cite{SplitVerb1,SplitVerb2}, namely that
the bosonic and fermionic semi-infinite Toda lattice 
hierarchies are connected by the supersymmetric partition function.

\subsection{Equivalence for the one Point Function}

The QCD partition function (\ref{Zf}) with $N_f$ mass degenerate fermionic
quarks  
satisfies the Toda lattice equation \cite{Kharchev}
\be 
 \delta_x^2 \log Z_{N_f}(x) = 2 N_f x^2 
\frac {Z_{N_f+1}(x)Z_{N_f-1}(x)}{[Z_{N_f}(x)]^2}.
\label{toda-chgue}
\ee
(Note that this is obtained from (\ref{Toda1PF}), satisfied by the
supersymmetric partition function, setting $n=N_f$ and $N_b=0$.)
In the replica approach we would like to take the $N_f\to0$ limit of this
equation. Using the representation (\ref{ZNf}) of the partition function, 
which depends continuously on $N_f$, we get \cite{SplitVerb1}
\be 
\lim_{N_f\to0}\frac 1{N_f} \delta_x^2 \log Z_{N_f}(x) & = & \delta_x x G(x),
\label{toda-replica-lhs}
\\
\lim_{N_f\to0} 2 x^2 \frac {Z_{N_f+1}(x)Z_{N_f-1}(x)}{[Z_{N_f}(x)]^2}
 & = & 2 x^2 Z_{1}(x)Z_{-1}(x).
\label{toda-replica-rhs}
\ee
However, the Toda lattice equation (\ref{toda-chgue}) was established 
starting from 
(\ref{Zf}), which is only well defined for positive integer values of
$N_f$, and the boundary condition $Z_{N_f=0}(x) =1$.  
The Toda lattice equation (\ref{toda-chgue}) therefore is not necessarily
valid for (\ref{ZNf}) when $N_f$ is non-integer.  
However, as we will now show, equating (\ref{toda-replica-lhs}) to
(\ref{toda-replica-rhs}) is identical to quenching the Toda lattice
equation using the supersymmetric method, since the
supersymmetric partition function satisfies the Toda lattice 
equation. To this end we write (\ref{toda-chgue}) as
\be 
\frac 1{N_f} \delta_x^2 \log Z_{N_f}(x) =
2x^2 \frac {Z_{N_f+1}(x)Z_{N_f-1}(x)}{[Z_{N_f}(x)]^2}
= \lim_{y\to x} 2x^2 
\frac {Z_{N_f+2,-1}(x|y)Z_{N_f,-1}(x|y)}{[Z_{N_f+1,-1}(x|y)]^2}.
\ee
These equalities are valid for positive integer values of $N_f$.
However, the r.h.s. of the second equality is also well defined for
$N_f= 0$. If we take this as the definition of the $N_f \to 0$ limit
of the middle term of this equation we get $2x^2 Z_1(x) Z_{-1}(x)$ exactly
as in (\ref{toda-replica-rhs}), but now without ambiguities in the
interpretation of $Z_{-1}(x)$. Moreover, in
appendix A it is shown that the supersymmetric partition function
entering the r.h.s satisfies the same Toda lattice equation 
\be 
\delta_x(\delta_x+\delta_y)\log Z_{N_f+1,-1}(x|y)
=2(N_f+1)x^2\frac {Z_{N_f+2,-1}(x|y)Z_{N_f,-1}(x|y)}{[Z_{N_f+1,-1}(x|y)]^2}.
\label{todaxy}
\ee
This equation is also well-defined for $N_f=0$ and in this case 
the limit $ y\to x$ is (cf. (\ref{22}))
\be
\lim_{y \to x} \delta_x(\delta_x+\delta_y)\log Z_{1,-1}(x|y)
&=& \delta_x x G(x).
\ee
This is precisely what we obtained in (\ref{toda-replica-lhs}) 
using the representation (\ref{ZNf}) of the fermionic partition
function. 
This shows that the replica limit of the Toda lattice equation
is necessarily   identical to the quenched limit of the
supersymmetric Toda lattice equation.

\subsection{Equivalence for the Two-Point Function}

We now consider the replica limit of the Toda lattice equation
affecting two indices simultaneously.
In \cite{SplitVerb2} it was shown that for integer $n >0$ 
\be
\label{Toda2pf-replica-1}
\del_{x_1}\del_{x_2}\log[Z_{n,n}(x_1,x_2)] =
4n^2 x_1x_2\frac{Z_{n+1,n+1}(x_1,x_2)Z_{n-1,n-1}(x_1,x_2)}{[Z_{n,n}(x_1,x_2)]^2}.
\ee
Making the analytic continuation in $n$ it was assumed \cite{SplitVerb2}
that the limit $n\to0$ connects this semi-infinite hierarchy to the 
semi-infinite hierarchy for $Z_{-n,-n}$, viz.
\be
\lim_{n\to0}\frac{1}{n^2}\del_{x_1}\del_{x_2}\log[Z_{n,n}(x_1,x_2)] =
4x_1x_2 Z_{1,1}(x_1,x_2)Z_{-1,-1}(x_1,x_2).
\ee
While the exact agreement of
the resulting quenched two-point correlation function 
with the known result \cite{TV-two} vindicates this replica limit, the 
underlying assumption again could not be proved.

We now prove the assumption for the two-point function using similar
arguments as for the one-point function.
Again we establish the 
troublesome connection between positive and negative integer values of
$n$ by quenching determinants as in the supersymmetric method. 
The ratio of partition functions in the r.h.s. of
(\ref{Toda2pf-replica-1}) can be written as
\be
\frac{Z_{n+1,n+1}(x_1,x_2)Z_{n-1,n-1}(x_1,x_2)}{[Z_{n,n}(x_1,x_2)]^2}
=\lim_{y_1\to x_1,y_2\to
  x_2}\frac{Z_{n+2,n+2,-1,-1}(x_1,x_2|y_1,y_2)Z_{n,n,-1,-1}(x_1,x_2|y_1,y_2)}
{[Z_{n+1,n+1,-1,-1}(x_1,x_2|y_1,y_2)]^2}.
\ee
As discussed in section \ref{subsec:TodaLattEq2pf} the ratio on the
r.h.s. satisfies a Toda recursion relation even before taking the limit of
degenerate masses 
\be
4(n+1)^2x_1 x_2\frac{Z_{n+2,n+2,-1,-1}(x_1,x_2|y_1,y_2)Z_{n,n,-1,-1}(x_1,x_2|y_1,y_2)}
{[Z_{n+1,n+1,-1,-1}(x_1,x_2|y_1,y_2)]^2} 
= \del_{x_1}\del_{x_2}\log[Z_{n+1,n+1,-1,-1}(x_1,x_2|y_1,y_2)]
\ee
In this relation we can safely replace $n$ by zero so that
\be
\label{Toda2pf-replica-2}
\lim_{n\to0}\frac{1}{n^2}\del_{x_1}\del_{x_2}\log[Z_{n,n}(x_1,x_2)] & = &
\lim_{y_1\to x_1,y_2\to
  x_2}\del_{x_1}\del_{x_2}\log[Z_{1,1,-1,-1}(x_1,x_2|y_1,y_2)] .
\ee
This shows that the replica limit of the Toda lattice equation
necessarily gives the same answer as the supersymmetrically quenched 
Toda lattice equation, thus justifying the analytic continuation in $n$.

\section{Conclusion}
\label{sec:conc}

The QCD partition function in the $\epsilon$ regime is equivalent 
to the microscopic limit of a chiral random matrix theory with
the same global symmetries. This equivalence holds for any number
of bosonic and fermionic quarks, and each partition function in this
family can be expressed as a determinant of Bessel functions. 
In this paper we have shown
that this family of partition functions satisfies a Toda lattice
equation which we have called the graded Toda lattice equation. 
Based on this result we have shown that the semi-infinite
hierarchy of bosonic or fermionic partition functions can be continued
to the limit of zero flavors which is given by the supersymmetric
partition function. 
This was the central assumption when taking the replica limit of the Toda
lattice equation. By proving this assumption we have thus shown the equivalence
of the replica limit of the Toda lattice equation and the supersymmetric 
method. 
Quenching the partition functions in 
the graded Toda lattice equation 
shows that the resolvent of the supersymmetric partition
function factorizes into the product of a bosonic and a fermionic partition
function.

In its simplest form, the graded
Toda lattice equation relates derivatives of the supersymmetric 
partition function to partition functions with one more flavor
or with one flavor less. This case is relevant for the analysis of
the one-point function and has been proved in all its generality.
For the analysis of the two-point function we need a graded Toda lattice
equation in which derivatives of the partition function are related
to partition functions with two more flavors or two less flavors. 
In addition  to the cases that were required for our
analysis, this Toda lattice equation was proved for several other
cases with a fixed number of flavors. 
Based on these results and known  
results for exclusively fermionic flavors, we have conjectured that
this equation can be generalized to an arbitrary number of 
bosonic and fermionic flavors. 
\vspace{2mm}

It is not yet clear whether variants of the Toda lattice equation 
are suitable for the
investigation of the replica limit of  real symmetric ($\beta_D=1$) and
quaternion real  ($\beta_D=4$) random matrix ensembles. However,
our work can be extended in several other directions.
First, our results can be extended in a straightforward way to
partition functions with a nonzero topological charge. The reason that
we considered only zero topological charge in this paper is that
it somewhat simplifies our notation. Second, in the formulation of
Kanzieper, the replica limit is taken of the Painlev\'e equation rather
than the Toda lattice equations. 
It would be very interesting to understand if the 
supersymmetric partition functions can also be incorporated in this
framework.  Third,
the spectral density of the quenched Dirac operator at non-zero
chemical potential in the limit of weak non-hermiticity was obtained
from the replica limit of a Toda lattice equation.
We expect that the supersymmetric generating
functional at non-zero chemical potential will satisfy a similar Toda lattice
equation so that the replica limit can also be justified in this case.
Fourth, the two-point function of the unitary ensemble  can also be obtained
from the replica limit of a Toda lattice equation. Also in this case we
believe that a generalization of the Toda lattice equation to supersymmetric
partition functions is possible. 
Finally, we have no doubt that a similar arguments can be applied 
to QCD in 3 dimensions.


\vspace{1cm}

\noindent
{\bf Acknowledgments:} We wish to thank Eugene Kanzieper for useful
correspondence on Painlev\'e equations and Poul Henrik Damgaard for
discussions of the replica trick. This work was supported in part by 
U.S. DOE Grant No. DE-FG-88ER40388.

\renewcommand{\thesection}{Appendix \Alph{section}}
\setcounter{section}{0}

\section{The graded Toda lattice equation where one index is affected} 
\label{App:Toda1PF}

In this appendix we prove the graded Toda lattice equation (\ref{Toda1PF})
relevant  
for the partially quenched one point functions, that is the relation where
one flavor is added and subtracted.
We start by slightly rewriting the general
expression for the supersymmetric partition function in (\ref{Z-nm})
\cite{FS,FSgen,FA,SplitVerb1} 
\be
\label{Z-nm-app1}
Z_{N_f,-N_b}(\{x_f\}|\{y_b\})=\frac{\det[\delta_{z_i}^{j-1}{\cal
J}(z_i)]_{i,j=1,..,N_f+N_b}}
{\prod_{j>i=1}^{N_f}(x_j^2-x_i^2)\prod_{j>i=1}^{N_b}(y_j^2-y_i^2)}.  
\ee
Here, $z_i=x_i$ for $i=1,\ldots,N_f$, and $z_{N_f+i}=y_i$ for $i=1,\ldots,N_b$.
Furthermore, the symbol ${\cal J}(z_i)\equiv I_0(x_i)$ for $i=1,\ldots,N_f$
and ${\cal J}(z_{N_f+i})\equiv K_0(y_i)$ for $i=1,\ldots,N_b$.  
In order to keep track of the
indices let us consider the last two fermionic flavors as mass degenerate,
that is $x_{N_f}=x_{N_f+1}$. In this case the partition function is 
equal to
\be
\label{ZandA}
Z_{N_f+1,-N_b}(x_{N_f},\{x_f\}_{f=1,..,N_f}|\{y_b\})
=\frac{\det[A_{N_f+1,-N_b}(x_{N_f},\{x_f\}_{f=1,..,N_f}|\{y_b\})]}
{\prod_{j>i=1}^{N_f}(x_j^2-x_i^2)\prod_{i=1}^{N_f-1}(x_{N_f}^2-x_i^2)2x_{N_f}
\prod_{j>i=1}^{N_b}(y_j^2-y_i^2)},\ee
where we have introduced the $(N_f+1+N_b)\times (N_f+1+N_b)$ matrix 
\be 
\label{defA} 
A_{N_f+1,-N_b}(x_{N_f},\{x_f\}_{f=1,..,N_f}|\{y_b\})
\equiv\left(\begin{array}{cccc}
I_0(x_1)& \delta_{x_1}I_0(x_1) & \cdots &  \delta_{x_1}^{N_f+N_b}I_0(x_1) \\
\vdots  &                      &        &         \vdots                    \\
I_0(x_{N_f})&            & \cdots & \delta_{x_{N_f}}^{N_f+N_b}I_0(x_{N_f})\\
\del_{x_{N_f}}I_0(x_{N_f})&            & \cdots & 
\del_{x_{N_f}}\delta_{x_{N_f}}^{N_f+N_b}I_0(x_{N_f})\\
K_0(y_1)& \delta_{y_1}K_0(y_1) & \cdots &  \delta_{y_1}^{N_f+N_b}K_0(y_1) \\
\vdots  &                      &        &         \vdots                    \\
K_0(y_{N_b})&            & \cdots & \delta_{y_{N_b}}^{N_f+N_b}K_0(y_{N_b})\\
\end{array}\right).
\ee 
The proof of the graded Toda lattice equation follows from the Sylvester
identity \cite{Sylvester} valid for the determinant of any matrix  
\be
\label{SylvEq1}
C_{i,j} C_{p,q} - C_{i,q}C_{p,j} = \det(A) C_{i,j;p,q},
\ee
where $C_{i,j}$ is the cofactor of matrix element $ij$ 
\be
C_{i,j}\equiv \frac{\del \det(A)}{\del A_{ij}},
\ee
and $C_{i,j;p,q}$ is the double cofactor of matrix elements $ij$ and
$pq$
\be
C_{i,j;p,q}\equiv \frac{\del^2 \det(A)}{\del A_{ij}\del A_{pq}}.
\ee
Setting $i=N_f+1$, $j=N_f+1+N_b$, $p=N_f$, and $q=N_f+N_b$ we have
\be
\label{SylvEq2}
C_{N_f+1,N_f+1+N_b} C_{N_f,N_f+N_b} - C_{N_f+1,N_f+N_b}C_{N_f,N_f+1+N_b} =
\det(A_{N_f+1,-N_b}) C_{N_f+1,N_f+1+N_b;N_f,N_f+N_b}.
\ee 
From the structure (\ref{defA}) of the matrix $A_{N_f+1,N_b}$ it follows that 
($\delta_{N_f+N_b}$ is defined in (\ref{def-deltaNfNb})) 
\be
C_{N_f+1,N_f+1+N_b}&=&\det A_{N_f,-N_b} , \nn \\
C_{N_f,N_f+N_b}&=& \frac{1}{x_{N_f}}\delta_{x_{N_f}}
\delta_{N_f+N_b} \det A_{N_f,-N_b},   \nn \\   
C_{N_f+1,N_f+N_b}&=& \delta_{N_f+N_b} \det A_{N_f,-N_b},  \\   
C_{N_f,N_f+1+N_b}&=& \frac{1}{x_{N_f}}\delta_{x_{N_f}} \det A_{N_f,-N_b},
\nn \\   
C_{N_f+1,N_f+1+N_b;N_f,N_f+N_b}&=& \nn \det A_{N_f-1,-N_b}.\nn
\ee
For completeness we write out 
\be
A_{N_f,-N_b}(\{x_f\}|\{y_b\})\equiv\left(\begin{array}{cccc}
I_0(x_1)& \delta_{x_1}I_0(x_1) & \cdots &  \delta_{x_1}^{N_f-1+N_b}I_0(x_1) \\
\vdots  &                      &        &         \vdots                    \\
I_0(x_{N_f})&            & \cdots & \delta_{x_{N_f}}^{N_f-1+N_b}I_0(x_{N_f})\\
K_0(y_1)& \delta_{y_1}K_0(y_1) & \cdots &  \delta_{y_1}^{N_f-1+N_b}K_0(y_1) \\
\vdots  &                      &        &         \vdots                    \\
K_0(y_{N_b})&            & \cdots & \delta_{y_{N_b}}^{N_f-1+N_b}K_0(y_{N_b})\\
\end{array}\right).
\ee
Inserting this in (\ref{SylvEq2}) we get
\be 
[\det A_{N_f,-N_b}]^2\delta_{x_{N_f}}\delta_{N_f+N_b} \log[\det A_{N_f,-N_b}]
= x_{N_f}\det A_{N_f+1,-N_b}\det A_{N_f-1,-N_b}.
\ee
Finally, using the relation (\ref{ZandA}) between the partition functions and
the matrices $A$ we obtain the desired graded Toda lattice equation
\be
\label{Toda1PF-app}
\delta_{N_f+N_b}\delta_{x_i}\log Z_{N_f,-N_b}(\{x_f\}|\{y_b\})
= 2 x_i^2\frac{ Z_{N_f+1,-N_b}(x_i,\{x_f\}|\{y_b\})
Z_{N_f-1,-N_b}(\{x_f\}_{f\neq i}|\{y_b\})}{[Z_{N_f,-N_b}(\{x_k\}|\{y_b\})]^2}.
\ee
Note that the Vandermonde determinants, $\Delta(\{z_i^2\})$, in the squared
masses do not contribute to the left hand side of the Toda lattice equation
since $\delta_{N_f+N_b}\log[\Delta(\{z_i^2\})]$ is a number. 

\vspace{5mm}

In the case where $n$ flavors have mass $x_{N_f}$ the proof is
similar. The partition function is now given by
\be 
\label{ZandA-n}
Z_{n,N_f-1,-N_b}(x_{N_f},\{x_f\}_{f\neq N_f}|\{y_b\})
=\frac{\det[A_{n,N_f-1,-N_b}(x_{N_f},\{x_f\}_{f\neq N_f}|\{y_b\})]}
{\prod_{j>i=1}^{N_f}(x_j^2-x_i^2)
\prod_{i=1}^{N_f-1}(x_{N_f}^2-x_i^2)^{n-1}(2x_{N_f})^{n(n-1)/2}
 \prod_{j>i=1}^{N_b}(y_j^2-y_i^2)},\ee
where we have introduced the $(N_f+n-1+N_b)\times (N_f+n-1+N_b)$ matrix 
\be
\label{defA-n}
A_{n,N_f-1,-N_b}(x_{N_f},\{x_f\}_{f\neq N_f}|\{y_b\})\equiv\left(\begin{array}{cccc}
I_0(x_1)& \delta_{x_1}I_0(x_1) & \cdots &
\delta_{x_1}^{N_f+n-2+N_b}I_0(x_1) \\
\vdots  &  &  &    \vdots        \\
I_0(x_{N_f})&  & \cdots & \delta_{x_{N_f}}^{N_f+n-2+N_b}I_0(x_{N_f})\\
\frac{1}{1!}\del_{x_{N_f}}I_0(x_{N_f})& & \cdots & \frac{1}{1!}\del_{x_{N_f}}\delta_{x_{N_f}}^{N_f+n-2+N_b}I_0(x_{N_f})\\
\vdots  &  &  &    \vdots        \\
\frac{1}{(n-1)!}\del_{x_{N_f}}^{n-1}I_0(x_{N_f})& & \cdots & \frac{1}{(n-1)!}\del_{x_{N_f}}^{n-1}\delta_{x_{N_f}}^{N_f+n-2+N_b}I_0(x_{N_f})\\
K_0(y_1)& \delta_{y_1}K_0(y_1) & \cdots &  \delta_{y_1}^{N_f+n-2+N_b}K_0(y_1) \\
\vdots  &                      &        &         \vdots                    \\
K_0(y_{N_b})&            & \cdots & \delta_{y_{N_b}}^{N_f+n-2+N_b}K_0(y_{N_b})\\
\end{array}\right).
\ee 
Using again the Sylvester identity, (\ref{SylvEq1}), this time with
$i=N_f+n-1$, $j=N_f+n-1+N_b$, $p=N_f+n-2$, and $q=N_f+n-2+N_b$ we find
the graded Toda lattice equation
\be 
\label{Toda1PF-app-n}
&& \hspace{-3cm }\delta_{N_f+N_b}\delta_{x_i}\log Z_{n,N_f-1,-N_b}(x_i,\{x_f\}_{f\neq
  i}|\{y_b\}) \\
&\hspace{1cm } = & 2 n x_i^2\frac{ Z_{n+1,N_f-1,-N_b}(x_i,\{x_f\}_{f\neq i}|\{y_b\})
Z_{n-1,N_f-1,-N_b}(x_i,\{x_f\}_{f\neq i}|\{y_b\})}{[Z_{n,N_f-1,-N_b}(x_i,\{x_k\}_{f\neq i}|\{y_b\})]^2}.\nn
\ee

\section{The graded Toda lattice equation where two indices are affected} 
\label{App:Toda2PF}

In this appendix we consider the graded Toda lattice equations
(\ref{TheToda2f-conj}) and (\ref{TheToda1f1b}) relevant for the two-point
functions. Although we 
have been able to check these equations in a number of cases
including those relevant for quenching the Toda lattice equation we do not
have a general proof.

The simplest possible example is 
\be 
\del_x\del_y\log Z_{1,-1}(x|y) = - 4 x y \frac{Z_2(x)Z_{-2}(y)}
                                          {[Z_{1,-1}(x|y)]^2}. 
\label{SimpleToda}
\ee
This result follows by rewriting the l.h.s. of this equation as
\be
[Z_{1,-1}(x|y)]^2\del_x\del_y\log Z_{1,-1}(x|y) =\det\left(
\begin{array}{cc}
Z_{1,-1}(x|y) & \del_x Z_{1,-1}(x|y) \\
\del_y Z_{1,-1}(x|y) & \del_y\del_x Z_{1,-1}(x|y) 
\end{array}\right),
\ee
and using the identity
\be
\left(
\begin{array}{cc}
Z_{1,-1}(x|y) & \del_x Z_{1,-1}(x|y) \\
\del_y Z_{1,-1}(x|y) & \del_y\del_x Z_{1,-1}(x|y) 
\end{array}\right) 
= \frac{1}{2} 
\left(
\begin{array}{cc}
Z_1(x) & \del_x Z_{1}(x) \\
\del_x Z_{1}(x) & \del_x^2 Z_{1}(x) 
\end{array}\right) 
\left(
\begin{array}{cc}
0&1\\1&0
\end{array}\right) 
\left(
\begin{array}{cc}
Z_{-1}(y) & \del_y Z_{-1}(y) \\
\del_y Z_{-1}(y) & \del_y^2 Z_{-1}(y) 
\end{array}\right) ,
\ee
which valid if the partition function is of the form
\be 
Z_{1,-1}(x|y) 
 = \frac 12[ Z_1(x) \delta_y Z_{-1}(y) + \delta_x Z_1(x) Z_{-1}(y) ].
\ee
This structure of the supersymmetric partition function easily follows by
integrating the bosonic variables before the Grassmann variables,
\be
 Z_{1,-1}(x|y) &=&
\int d \alpha d\beta 
I_0(x(1+ \alpha\beta/2)K_0(y(1-\alpha\beta/2)\nonumber \\
       &= &\frac 12[ I_0(x) \delta_y K_0(y) + \delta_x I_0(x) K_0(y) ].
\ee

{\bf The general case:} Our conjecture is that the generalization of the
above graded Toda lattice equation is given by
\be 
\label{TheToda1f1b-app}
\delta_{x_i}\delta_{y_j} \log Z_{N_f,-N_b}(\{x_f\}|\{y_b\}) 
 = -4nm x_i^2 y_j^2 \frac{ Z_{N_f+1,-N_b+1}(x_i,\{x_f\}|\{y_b\}_{b\neq j})
Z_{N_f-1,-N_b-1}(\{x_f\}_{f\neq i}|y_j,\{y_b\})}
{[Z_{N_f,-N_b}(\{x_f\}|\{y_b\})]^2},
\ee 
where $n$ is the degeneracy of $x_i$  and
$m$ is the degeneracy of $y_j$ in $Z_{N_f,-N_b}(\{x_f\}|\{y_b\})$,
respectively.
We have explicitly checked this equation for the cases listed in the table
below: 
\vspace{1cm}

\renewcommand{\arraystretch}{1.2}
\begin{center}
\begin{tabular}{|c|c|c|c|c|} 
\ Eq (\ref{TheToda1f1b-app}) \ & \ $N_f$ \ & \ $N_b$ \ & \ $n$ \ & \ $m$ \   \\
\hline
case 1 & 1 & 1 & 1 & 1  \\
case 2 & 2 & 2 & 1 & 1  \\
case 3 & 2 & 2 & 2 & 1 \\
case 4 & 2 & 2 & 2 & 2 \\
case 5 & 3 & 1 & 3 & 1 \\
case 6 & 1 & 3 & 1 & 3
\end{tabular}
\end{center}

\vspace{1cm}

A similar equation holds when both derivatives are with respect to 
fermionic masses. It is conjectured to be given by
\be
\label{TheToda2f-conj-app}
 \delta_{x_i}\delta_{x_j} \log Z_{N_f,-N_b}(\{x_f\}|\{y_b\}) 
 = 4 nm 
x_i^2 x_j^2 \frac{ Z_{N_f+2,-N_b}(x_i,x_j,\{x_f\}|\{y_b\})
Z_{N_f-2,-N_b}(\{x_f\}_{f\neq i,j}|\{y_b\})}
{[Z_{N_f,-N_b}(\{x_f\}|\{y_b\})]^2},
\ee
where $n$ is the degeneracy of $x_i$ 
and $m$ is the degeneracy of $x_j$ in $Z_{N_f,-N_b}(\{x_f\}|\{y_b\})$, in
this order. We have verified this identity for the following cases:

\renewcommand{\arraystretch}{1.2}
\begin{center}
\begin{tabular}{|c|c|c|c|c|} 
\ Eq (\ref{TheToda2f-conj-app}) \ & \ $N_f$ \ & \ $N_b$ \ & \ $n$ \ & \ $m$ \   \\
\hline
case 1 & any & 0 & any & any  \\
case 2 & 2 & 2 & 1 & 1  \\
case 3 & 3 & 1 & 2 & 1 \\
case 4 & 4 & 1 & 2 & 2 \\
case 5 & 4 & 1 & 3 & 1 
\end{tabular}
\end{center}

\end{document}